# Scaling and correlations in 3 bus-transport networks of China

Xinping Xu [1,*], Junhui Hu [2], Feng Liu[1], Lianshou Liu[1]

[1]*Institute of Particle Physics, Hua-Zhong Normal University, Wuhan 430079, China*

[2]*College of Physics and Information Technology, Guangxi Normal University, Guilin 541004, China*

Abstract: We report the statistical properties of 3 bus-transport networks (BTN) in three different cities of China. These networks are composed of a set of bus lines and stations serviced by these. Network properties including the degree distribution, clustering and average path length are studied in the space L and P respectively. It is demonstrated that the degree in the space L obeys a power law distribution while that in the space P follows an exponential distribution. Degree of a given station in the space P and number of lines that service that station have linear positive correlations. A small average path length and large clustering coefficient is also presented in the space P. Besides, we create a weighted network representation for BTN with lines mapped to nodes and number of common stations to weights between lines. In such representation, both the cumulative degree and strength distributions obey two-regime power laws with different exponents while the cumulative distributions of weights have power-law tails. A power-law behavior between strength and degree, $s(k) \sim k^b$, with a value of $b$ close to 1.1 is also observed.

## 1. Introduction

During the last few years several transport networks have already been investigated using various concepts of statistical physics of complex networks [1]. These networks are called public transport networks, in which the nodes are the stations of public transport and the edges are the links connecting them along the route (see Fig.1). These public transport networks can be classified as railway networks [2,3], airport networks [4－9], subway (underground) transport networks [10－12] and bus-transport networks [13,14] etc. As other type of networks, all these networks display the small world behavior and scale free structure.

In this paper, we investigate 3 bus-transport networks (BTN) located in 3 different cities of China. In BTN, the nodes are the bus stations and the edges are the bus lines connecting them along the route. In order to obtain general statistical properties of a network structure, we choose 3 big cities which have a large numbers of routes and stops. The results presented below are based on an analysis of bus-transport networks of Beijing (516 routes and 3938 stations), Shanghai (501 routes and 2063 stations) and Nanjing (174 routes and 1150 stations). The data of bus-transport for the above cities were downloaded from the Internet. Compared to other types of networks, BTN appear extraordinary and unique due to the following features: 1) there are two topological representations for BTN, the so-called space L and space P. Basing on the concepts of space L and space P, we will show that scaling laws may govern intrinsic features of the physical quantities, 2) large differences in network size. Number of stations varies from a few hundred to a few thousand, 3) a sequence of stations is joined by more than one line. This is the familiar situation when one can go from one station to another by different bus lines without making a change.

The paper is structured as follows: In the next section, we give a brief description of the concepts for the space L and space P. In section 3, we will show that the degree in the space L obeys a power law distribution while the cumulative degree in the space P follows an exponential distribution. Section 4 analyzes the clustering and average path length. In section 5, we introduce another representation to describe BTN and explore more scaling laws for BTN. Discussion and conclusions

## 2. The concepts of space L and P

Sen et al. [2] have introduced a new topology describing the public transport networks－the idea of the space P which indicating two arbitrary stations are connected by a link when there is at least one bus which stops at both the stations. Generally, we can present the public transport network in two different topological representations. The first topological representation is the space L which consists of nodes being bus stops provided a link between two nodes exists. The distance in such a space is measured by the total number of stops passed on the shortest path between two nodes. The node degree k in this topology is just the number of directions one can take from a given stops. The second representation is the space P, in which an edge is formed between two nodes given that there is a bus traveling between them. It is obvious that in the space P the distances are numbers of transfers (plus one) needed during the travel and distances are much shorter than in the space L. Both spaces are presented at Fig.1.

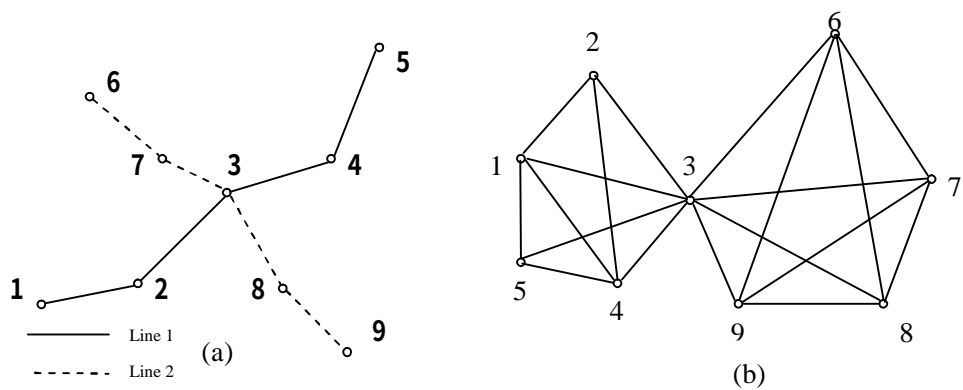

Fig.1. Illustration of the space L (a) and the space P (b).

## 3. Degree distribution

First, we examine the number of buses (or routes) that service a certain station. The number of buses that pass a given station reflects the convenience of the station when one travels to other stations. Fig.2 (a) shows the distribution of the number of lines (or routes) that services each station. It is obvious that the number of buses that pass through a given station follows a power law distribution. Such a distribution may be explained as the following. There are a large number of suburban stations, and the stations in suburb usually have fewer buses that traveling through them compared to the stations in the central city. Fig.2 (b) shows the degree distribution in the space L. Since $k=1$ nodes are ends of transport routes, the number of nodes of degree $k=1$ is smaller as compared to the number of nodes of degree $k=2$, therefore points $k=1$ of the degree distribution in space L are peculiar. The maximal probability observed for nodes with degree $k=2$ means that a typical stop is directly connected to two other stops. Some bus stations can have a relatively high degree value (in some cases above 20) but the number of such stations is very small. The degree distribution in the space L can be described by power-laws. The exponents of these power laws in 3 cities are in the range of (2.5, 3.0).

In order to reduce large fluctuations, we use the cumulative distribution $P(k)$ [15] to describe the degree distribution in the space P. The cumulative distribution in the space P for 3 cities can be presented as an exponential function,

$$P(k) = Ae^{-ak} \tag{1}$$

as shown in Fig.3 (a). In Fig.3 (b), we plot the degree in the space P as a function of the number of lines that services a given station, which implies that stations having more buses passed through has a large degree in space P. Such strong correlation can be understood as follows. Consider a given station $i$ as a crossing in several lines, suppose each line has equal number of stops and there are no other overlapped stops in the lines, the degree in space P is given by

$$k_p = (a-1)n \qquad (2)$$

where $a$ is the number of stops of each line, and $n$ is the number of lines that services station $i$. Therefore, the degree of station $i$ in space P is proportional to the number of lines that services station $i$.

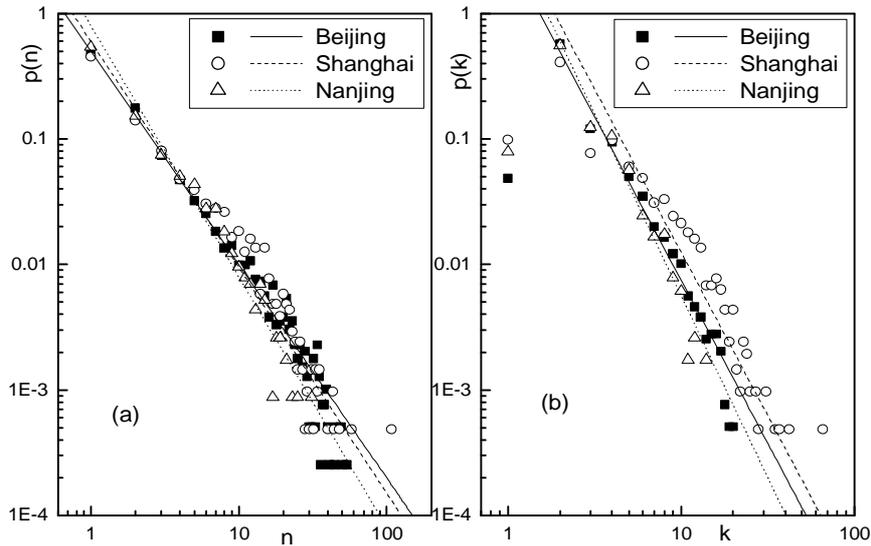

Fig.2. (a) Distribution for the number of lines that service each station. (b) Degree distribution in the space L.

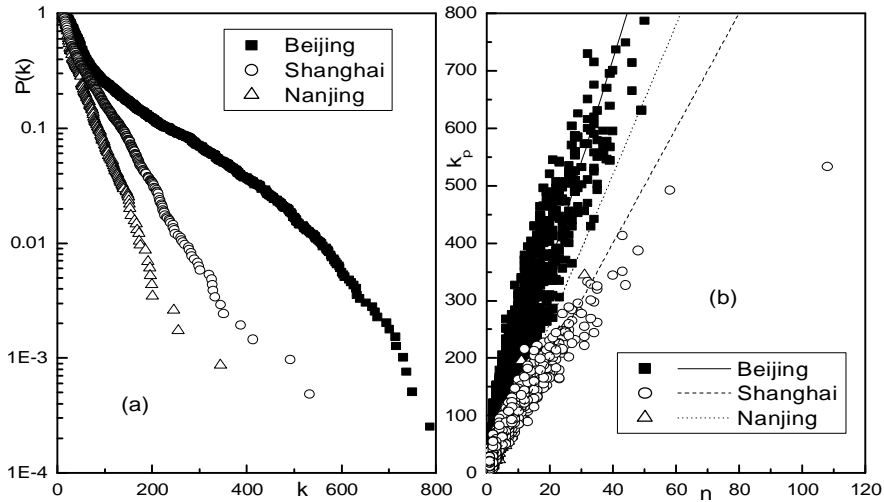

Fig.3. (a) Cumulative degree distribution in the space P. (b) Degree in space P versus the number of lines that services each station. The lines are least-squares fits to the form $k_p \sim n$.

## 4. Clustering and average path length

In this section, we consider the clustering and average path length. First, we check the clustering spectrum, or the degree-dependent local clustering, which can be described by the clustering coefficient as a function of degree. In many instances, the clustering spectrum exhibits also a power-law behavior, $C(k) \sim k^{-a}$. A value of $a$ close to 1 has been empirically observed in several real networks, and analytically found in some growing network models [18−20]. We only report properties of the clustering spectrum in the space P since the data in the space L are meaningless. The clustering spectrum in the space P is shown in Fig.4. The clustering coefficient C is independent of k in the region (0, 15). As k grows, C decreases as $C(k) \sim k^{-a}$ with values of exponents $a$ close to 0.7. The nontrivial clustering may indicate a hierarchical and modular structure in bus-transport networks.

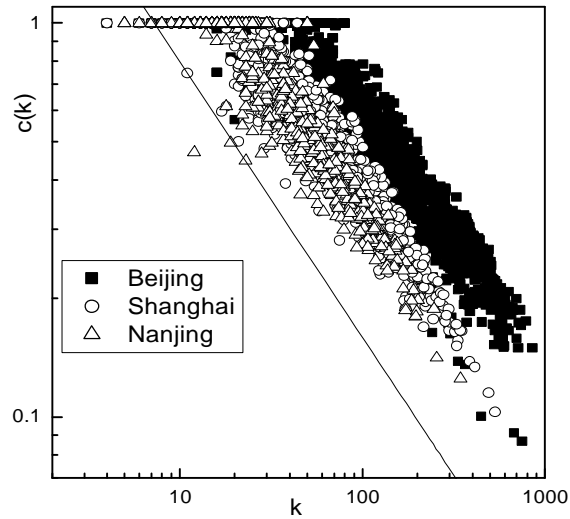

Fig.4. Clustering spectrum in space P. The solid line is a power law $C(k) \sim k^{-0.7}$.

The average clustering coefficient and average path length in the space L and P are listed in TABLE.I. The average path length in the space P does not exceed 3 in all the 3 cities, which means that in order to travel between two different stations one needs in average no more than two transfers. Other values of the parameters including average degree and assortativity coefficient defined by Newman in the space L and P are also listed in Table I.

## 5. New network representation for BTN

Next, we introduce another network representation for BTN. We regard each line (or route) as a node; two nodes have interaction if there are intersections between these two nodes. In this perspective, BTN can be symbolized by a symmetrical weight matrix $W$ whose element $W_{ij}$ (or $W_{ji}$) is the number of common stations between lines $i$ and $j$. Hence, the degree and strength of vertex $i$, $k_i$ and $s_i$ can be given by

$$k_i = \sum_{j \neq i} h(W_{ij} - 1) \qquad (3)$$

and

$$s_i = \sum_{i \neq j} W_{ij} \qquad (4)$$

where $h(x)$ is a unit step function, which takes the value 1 for $x \geq 0$ and 0 otherwise.

First we consider the distributions of $k_i$ and $s_i$ respectively. Here the cumulative distribution, widely used in economies and well known as the Pareto law [21], is adopted to reduce the statistical errors arising from the limited system size. The cumulative form, $P(k_i > k)$ (or $P(s_i > s)$) gives the probability that a given line $i$ has a degree (or strength) larger than $k$ (or s). Figure 5(a) and (b) present behaviors of the two distributions in three different cities. It is amazing to find that all two distributions follow a two-regime power law with two different exponents, known as double Pareto law [22], with a turning point at degree value $k_c$, which can be well prescribed by the following expression:

$$P(K > k) \sim \begin{cases} k^{-g_1}, & for \quad k \leq k_c \\ k^{-g_2}, & for \quad k > k_c \end{cases} \quad (5)$$

where $g_1$ and $g_2$ are the respective degree exponents of two separate power laws. By means of fitting, exponents pairs ($g_1, g_2$) of the degree and strength distributions in Fig. 5 are given in Table II.

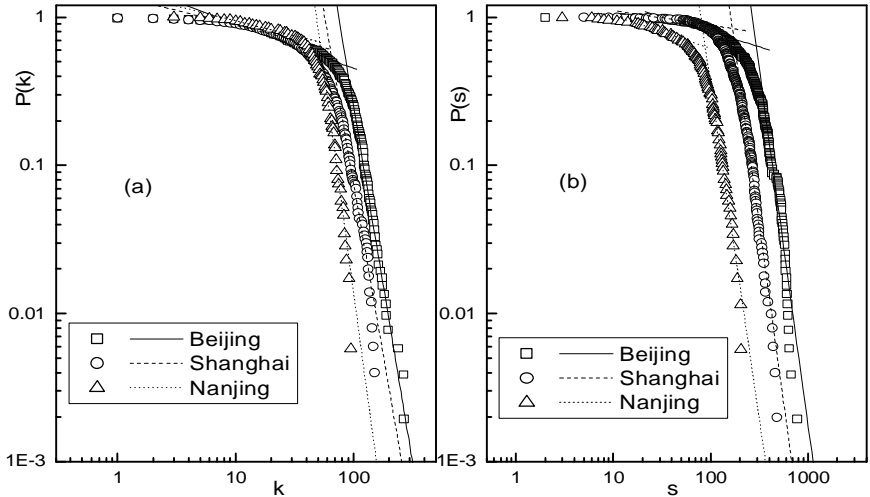

Fig.5. Cumulative degree (a) and strength (b) distributions of BTN in Beijing (squares), Shanghai (circles) and Nanjing(triangles).

A rather particular feature of BTN is that many routes service common subsets of stations. In order to explore this characteristic, we study the distribution of the common stations shared by two arbitrary lines. We find that the cumulative distribution of $W_{ij}$ can be described by a power-law

$$P(W_{ij} > W) \sim W^{-g} \quad (6)$$

as shown in Fig.6(a). Through a simple algebra, we can get the original distribution $P(W) \sim W^{-(g+1)}$. Such a power-law behavior indicates that the probability of finding a long sequence of stations shared by two lines is nonzero.

To shed some light on the relationship between the node strength and degree, we investigate the dependence of $s_i$ on $k_i$. We find that the strength $s(k)$ of nodes with degree $k$ increases with the degree as

$$s(k) \sim k^b. \qquad (7)$$

Fig.6 (b) is a plot of $s(k)$ versus $k$, which gives a value of $b$ close to 1.1. This value implies that the strength grows faster than their degree. Similar result is also observed in Ref.[7]. More features of our weighted network, including the average degree <k> and strength <s>, are given in TABLE II.

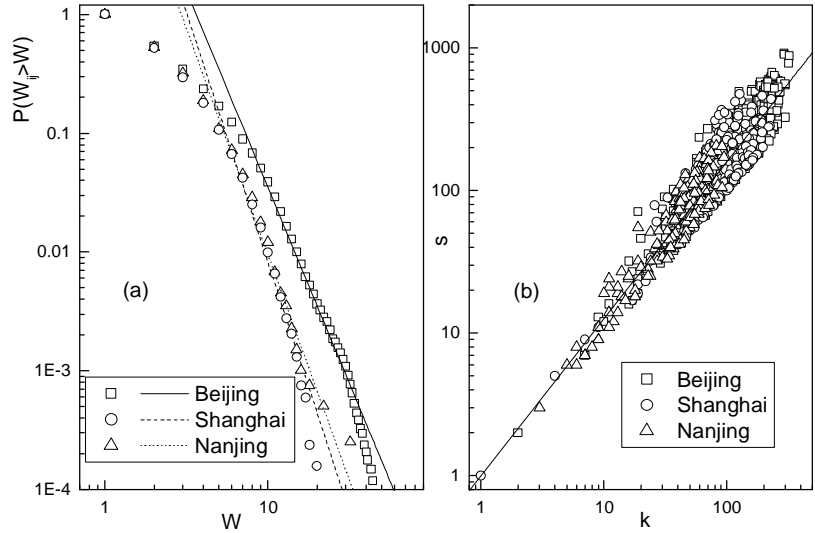

Fig.6 (a) Cumulative weight distributions. (b) Strength $s$ as function of the degree $k$ of BTN. The solid line in figure (b) indicates that $s$ versus $k$ in three cities follow nearly the same power law $s(k) \sim k^{1.1}$

## 6. Discussion and conclusion

In this paper, we have analyzed the statistical properties of 3 bus-transport networks in China. We have observed that these transport networks follow universal scalings. In the space L the degree distribution and distribution for the number of lines that services each station obey power laws while the cumulative degree distribution in space P follows an exponential distribution. Small world behavior is observed in both topologies but it is much more distinct in the space P and the hierarchical structure of network is also deduced from the behavior of $c(k)$. Furthermore, we regard BTN as a weighted network with lines mapped to nodes and number of common stations to weights between lines. Interestingly, both the degree and strength distributions in this representation follow a double Pareto law with different exponents. This gives similar results as in Ref. [8]. A heavy tailed power-law for the weight distribution and a power-law dependence between the strength and degree is also observed.

We would like to point out that the new network representation for BTN introduced by us, open a novel understanding for transport networks in the perspective of information handling. Since the traffic of transport network is constrained by the information involved in locating the destination, the actual traveling distance may be less restrictive than the information needed to locate the correct address in many cases. We assume that it does not cost any information to travel within a given route but cost additional information when people transfer from one line to another. In this sense, the new representation maps the BTN to an information city network [23], which present a generic way to estimate the information needed to navigate in a city. This should be an interesting topic and worth investigating.

**7. Acknowledgement** This work is supported by NSFC under projects 10375025, 10275027 and by the MOE under project CFKSTIP-704035.

TABLE I. Parameters of three BTN in the space L and P.

| Space and Networks / Parameters | Space L | | | Space P | | |
|---|---|---|---|---|---|---|
| | Beijing | Shanghai | Nanjing | Beijing | Shanghai | Nanjing |
| Average degree | 3.22 | 4.59 | 2.88 | 94.19 | 55.78 | 41.06 |
| Average clustering coefficient | 0.15 | 0.21 | 0.09 | 0.77 | 0.73 | 0.78 |
| Assortativity coefficient | 0.16 | 0.05 | 0.11 | 0.04 | -0.04 | 0.06 |
| Average path length | 12.56 | 7.13 | 12.42 | 2.54 | 2.59 | 2.66 |

TABLE II. Comparison of relevant variables in three cities, where $g_1$ and $g_2$ are exponents of two power laws of cumulative degree and strength (in parentheses) distributions, $g$ is the exponent of the cumulative weight distributions, <k> the average degree; <s> the average strength.

| Cities | Beijing | Shanghai | Nanjing |
|---|---|---|---|
| $g_1$ | 0.32(0.30) | 0.21(0.19) | 0.12(0.11) |
| $g_2$ | 4.80(4.76) | 4.62(4.56) | 5.88(5.62) |
| $g$ | 3.3 | 4.2 | 3.8 |
| <k> | 131.2 | 100.2 | 45.5 |
| <s> | 249.4 | 164.4 | 76.1 |

*Corresponding author. E-mail address: xuxp@ihep.ac.cn